\documentclass[aps,prb,reprint,floatfix,superscriptaddress,showpacs,letterpaper,longbibliography]{revtex4-1}

\usepackage{graphicx}
\usepackage{bm}
\usepackage{amsmath}
\usepackage{amsfonts}
\usepackage{epstopdf}
\usepackage{xfrac}

\bibpunct{[}{]}{,}{n}{}{}

\newcommand{\LNO}{LiNO$_3$}
\newcommand{\NO}{NO$_3^{-}$}

\newcommand{\hBN}{{\it h-}BN}

\newcommand{\abi}{{\sc abinit}}

\begin{document}

\title{Resonant X-ray Emission and Valence-band Lifetime Broadening in LiNO$_3$}

\author{John Vinson}
\affiliation{National Institute of Standards and Technology, 100 Bureau Drive, Gaithersburg, MD 20899}
%\affiliation{Material Measurement Laboratory, National Institute of Standards and Technology, Gaithersburg, MD 20899}

\author{Terrence Jach}
\affiliation{National Institute of Standards and Technology, 100 Bureau Drive, Gaithersburg, MD 20899}

\author{Matthias M\"{u}ller}
\affiliation{Physikalisch-Technische Bundesanstalt, Abbestra{\ss}e 2-12, 10587 Berlin, Germany}

\author{Rainer Unterumsberger}
\affiliation{Physikalisch-Technische Bundesanstalt, Abbestra{\ss}e 2-12, 10587 Berlin, Germany}

\author{Burkhard Beckhoff}
\affiliation{Physikalisch-Technische Bundesanstalt, Abbestra{\ss}e 2-12, 10587 Berlin, Germany}

\date{\today}

\begin{abstract}
X-ray absorption and resonant inelastic x-ray scattering measurements are carried out on lithium nitrate LiNO$_3$. 
The $\sigma$ orbitals around the nitrogen atoms exhibit a large lifetime effect. 
Experimentally, this is manifest as an apparent weakening of the x-ray emission signal from these states, but a closer examination shows that instead it is due to extreme broadening. 
This echos previous studies on ammonium nitrate, which, despite large differences in the cation and space group, showed a similar effect associated with the nitrate. 
Using first-principles {\it GW} self-energy and Bethe-Salpeter equation calculations we show that this effect is due in part to short quasi-hole lifetimes for the orbitals constituting the NO~$\sigma$ bonds.  

\end{abstract}

\maketitle

\section{Introduction}

Resonant inelastic x-ray scattering (RIXS) maintains the advantages of near-edge x-ray spectroscopy techniques and brings them to bear on low-energy excitations. 
Near-edge x-ray spectroscopy refers to techniques where the x-ray is tuned near a core-level resonance of the target element. 
The core-level resonances are well-separated throughout the periodic table, making these techniques element specific. 
Further, the localized nature of the core orbitals results in x-ray excitations that are also localized around the absorbing atom. 
Lastly, x-rays have longer penetration depths than optical probes, allowing the observation of bulk properties, buried interfaces, or samples under non-vacuum conditions. 
RIXS is a photon-in--photon-out spectroscopy, decoupling the x-ray energy from the energy transferred to the system, i.e., the energy of the excitations that are measured.
The types of phenomena accessible through RIXS are dependent on the energy loss that is measured. 
At the low end, starting at meV, are phonon, magnon, or charge-density wave excitations, and at higher energies through several tens of electron-Volts are electron-hole excitations that reveal details of the electronic structure. 

In this work we investigate the electronic structure of LiNO$_3$ using RIXS, focusing on details of the quasiparticles that make up the valence and conduction band states. 
Specifically, our RIXS measurements reveal the quasiparticle lifetimes of valence-band excitations. 
In the RIXS process, a valence exciton (conduction-electron--valence-hole pair) is considered to be the final state --- the excitation caused by the incident x-ray being absorbed and reemitted. 
This valence exciton, however, is itself an excitation and will decay, reflecting the lifetimes of its constituent electron and hole. 
A short lifetime implies an uncertain excitation energy and a broadening of the associated peak in the spectrum. 
By assumption, a quasiparticle excitation is long-lived with a lifetime broadening (the imaginary component of their energy) that is small compared to its energy measured from the Fermi level $\Gamma \ll \vert E - E_F\vert $, allowing it to be treated as a single-particle excitation \cite{Abrikosov}. 
%The assumption of quasiparticles is that single-particle excitations are long-lived with a lifetime broadening (the imaginary component of their energy) that is small compared to its energy measured from the Fermi level $\Gamma \ll \vert E - E_F\vert $  \cite{Abrikosov}. 
%%compared to the band gap in insulators: $\Gamma \ll E_g (E_F)$ \cite{Landau}. 
It was previously observed that in the case of ammonium nitrate, features assigned to the NO~$\sigma$ bonding orbitals had an anomalously large lifetime broadening \cite{PhysRevB.90.205207,PhysRevB.94.035163}. 
In the current work, interpretation of the nitrogen K-edge absorption and emission spectra is simplified by the presence of only a single nitrogen species. 
LiNO$_3$ forms a markedly different crystal structure from ammonium nitrate, but the NO~$\sigma$ orbitals show a similar lifetime broadening effect.

An experimental determination of valence lifetime broadening via RIXS is complicated by a variety of effects. 
The valence and conduction orbitals have energies and transition matrix elements that depend on details of the electronic structure. 
This situation is further complicated by the prevalence of vibrational disorder, including zero-point motion. 
The excitations are modified by interactions in both the intermediate state, following x-ray absorption, and in the final-state valence exciton. 
Therefore, we support our measurements of \LNO{} with first-principles modeling that explicitly includes these effects. 

Our paper continues by briefly discussing the structure of \LNO{} before presenting the phonon response and vibrational disorder. 
We then calculate the band structure and quasiparticle lifetimes for \LNO{}. 
Next, we present measured x-ray absorption and emission spectra and compare with calculations. 
We find that the calculations are able to reproduce the measured absorption spectrum and the main features of the x-ray emission. 
However, away from threshold the RIXS spectra broaden, and the main $\pi$ feature splits into two peaks. 
These changes, which we believe to believe to be due to structural relaxation during the RIXS process, are not captured in the calculations. 
Near threshold, the measured broadening of the NO~$\sigma$ emission feature is well-matched by our {\it GW} calculation of the imaginary component of the valence-band self-energy.

\section{Phonons and Quasiparticles}

\LNO{} crystalizes in the {\it R}$\bar{3}${\it c} (\#167) space group and is isostructural with calcite. 
In contrast, in our previous study we investigated ammonium nitrate which forms an orthorhombic structure \cite{PhysRevB.90.205207,PhysRevB.94.035163}. 
The ground-state properties of \LNO{} were calculated using density-functional theory (DFT) as implemented in the {\sc abinit} code \cite{abinit1,*abinit0}. 
The Perdew-Burke-Ernzerhof (PBE) exchange-correlation potential was used \cite{PhysRevLett.77.3865}, including semi-empirical dispersion corrections (DFT-D) \cite{Grimme}. 
The dispersion corrections were truncated at $2\times10^{-10}$~Ry for an effective pair radius of 97~a.u. 
Following \cite{PhysRevB.93.144304}, we neglect the three-body terms in both the structural optimization and later in the phonon calculations. 
Pseudopotentials were taken from the PseudoDojo collection \cite{pspdojo1,*pspdojo0} and generated using {\sc oncvpsp} \cite{PhysRevB.88.085117,*oncvp}. 
A planewave cutoff energy of 130~Ry was used, and the lattice parameters and atomic positions were adjusted until the forces on each atom were less than $5\times10^{-6}$~Ry~a.u.$^{-1}$.  
The calculated lattice constants and atomic positions are presented in Table~\ref{positions} alongside measured values \cite{doi:10.1021/ic00085a025}. 
(The thermal parameters B will be presented in section~\ref{phonons}.)  
In our calculations, a slight compression is found along the {\it c} axis (perpendicular to the \NO{} planes), and a negligible elongation along the {\it a} axis.
Overall, the agreement between our calculations and the experimental values are good. 
A number of previous calculations of the structure of \LNO{} have been carried out also using DFT-D \cite{KORABELNIKOV2015,YEDUKONDALU2016}. 
These prior calculations found comparable lattice constants, less than 1~\% larger, albeit using an earlier version of the DFT-D two-body corrections.

\begin{table}
 %\begin{center}
 \caption{ Calculated structure and isotropic thermal parameters (B) of \LNO{} compared to measured literature values \cite{doi:10.1021/ic00085a025}. 
 With the exception of the oxygen $x$ value, all of the atomic coordinates are constrained by the symmetry, denoted by a `-' symbol. 
 The calculations are numerically converged to beyond the most significant digit shown.   }
 \begin{tabular}{  c  c  c  c c c c}
\hline
& & {\it a} (nm)\;\,\,\, & & {\it c} (nm)\;\,\;\,\,\, \\
\hline
&exp  & 0.46920(3) && 1.52149(13)\\
&calc  & 0.47093\quad\,\, &&     1.50709\quad\;\,\,\, \\
\hline
\hline
atom & & x & y & z & B (a.u.$^2$)\\ 
\hline
Li & exp & 0 & 0 & 0 & 5.79(14)\\ 
 & calc & - & - & - & 6.22\quad\;\,\,\,\\ 
\hline
N & exp & 0 & 0 & \sfrac{1}{4} &2.88(3) \!\,\, \\
 & calc & - & - & - & 2.39\quad\;\,\,\,\\ 
\hline
O & exp & 0.2667(1) & 0 & \sfrac{1}{4}  & 4.00(4) \!\,\, \\ 
 & calc & 0.2678\quad\,\, & - & -  & 3.38\quad\;\,\,\, \\  
\hline
\end{tabular}
%\end{center}
\label{positions}
\end{table}

\subsection{Phonon bandstructure and disorder}
\label{phonons}

To account for the effects of vibrational disorder we calculate and average together x-ray spectra from an ensemble of disordered structures. 
Each structure in the ensemble is a possible configuration of a subset of the atoms within the crystal. 
To generate the ensemble of disordered structures we make use of the phonon-mode approach introduced previously \cite{PhysRevB.90.205207,PhysRevB.94.035163}, and expanded on in \cite{PhysRevB.96.205116} and separately in \cite{PhysRevB.92.144310}.
Briefly, the phonon modes of a structure are assumed to be harmonic. 
Within the harmonic approximation, each mode is an independent one-dimensional quantum harmonic oscillator (QHO). 
At any temperature, the probability distribution function of a QHO follows a normal distribution which can be written in terms of the natural length-scale $x$ and  dimensionless  $\tilde{T}$ which is the temperature $T$ rescaled by  Boltzman's constant $k_\text{B}$ and the frequency of the mode $\omega$
\begin{align}
\label{phononEq}
&P(x;\tilde{T}=k_\text{B}  T / \omega ) = \frac{ 1}{ \sqrt{2 \pi} u(\tilde{T})} e^{-x^2 / 2u^2(\tilde{T})} \\
& u^2(\tilde{T}) = \left[ \sfrac{1}2{} + n(\tilde{T}) \right]; \quad \quad n(\tilde{T}) = \left[ e^{\sfrac{1}{\tilde{T}}} - 1 \right]^{-1} \nonumber
\end{align}
The Bose statistics ensure that even as the occupation number $n$ for the QHO goes to zero at low temperatures, $k_B T \ll  \omega$, the mode continues to have a variance of $\sfrac{1}{2}$, i.e., zero-point disorder.
A disordered snapshot of the system is created by randomly generating displacements for each phonon mode according to Eq.~\ref{phononEq}. 
Each snapshot in the ensemble uses a different set of random displacements. 
The thermal parameters shown in Table~\ref{positions} are proportional to the isotropically averaged variances $u^2$ over all the included phonon modes $\lambda$ 
\begin{align}
B_i(T) &=  8 \pi^2 \sum_\lambda x_{i\lambda}^2 u^2(k_\beta T / \omega_\lambda) \\
 x_{i\lambda}^2 &= \left| \vec{\xi}_{i \lambda} \right|^2 \big[ M_i \omega_\lambda \big]^{-1} \nonumber
\end{align}
where $\vec{\xi}_{i \lambda}$ is the phonon eigenvector and $M_i$ is the mass of the $i$th atom. 
If the sum over phonon modes is done in reciprocal space, there will be an additional normalization by the number of included reciprocal space points. 
Non-isotropic thermal parameters can be determined by projecting $\vec{\xi}$ onto various unit vectors before taking the square.

\begin{figure}
\includegraphics[height=3.1in,trim=0 35 15 60,angle=270]{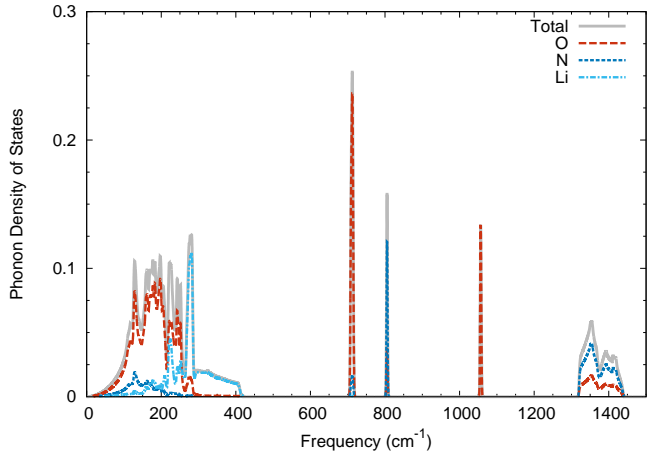}
\caption{The phonon density of states of \LNO{}. The lithium atoms only participate in the lower frequency modes, while the intra-ionic \NO{} modes $E$ bending, $A_{2}$, $A_{1}$, and $E$ stretching are at 716~cm$^{-1}$, 804~cm$^{-1}$, 1058~cm$^{-1}$, and from around 1300~cm$^{-1}$ to 1450~cm$^{-1}$, respectively. 
The small $\approx2$~cm$^{-1}$ splitting between the intra-ionic \NO{} gerade and ungerade modes indicates that the influence of the lithium ion is weak. }
\label{phononDOS}
\end{figure}

The phonon modes are calculated using density-functional perturbation theory \cite{DFPT} as implemented in the \abi{} code  \cite{abinit1,PhysRevB.93.144304}. 
The electron wave functions were determined on an $8^3$ $\Gamma$-centered mesh, and the interatomic force constants were determined on a $8^3$ mesh. 
The phonon density of states (DOS), shown in Fig.~\ref{phononDOS}, was constructed via Fourier interpolation \cite{PhysRevB.55.10355}.
The atom-projected contributions are also shown. The intra-ionic modes of the \NO{}, those for an isolated trigonal AB$_3$ molecule, are well-separated above 600~cm$^{-1}$. 
The calculated \NO{} modes agree with Raman measurements to within 3~\% \cite{doi:10.1063/1.1672185,BROOKER1978657}.
Note that at room temperature $298\,\text{K} \times k_\text{B} =207$~cm$^{-1}$. 
Therefore, the thermal occupation of the higher-frequency modes is negligible, and these modes are dominated by zero-point motion. 
The lower-energy modes are primarily rotation and libration of the \NO{} and counter motion of the Li$^{+}$ and \NO{} ions. 
These modes do not change the geometry of the \NO{} and have a small effect on the nitrogen {\it K}-edge spectra.

\subsection{Bandstructure and Self-energy}

Using the previously determined ground-state structure (neglecting vibrations), we have calculated the electronic structure using DFT with the same parameters as noted above, except that dispersion corrections are not included. 
The formulation of DFT-D does not effect the energy of the electronic part of the Hamiltonian, and is therefore used only for the determination of the structure and phonon modes. 
In addition to calculations of the electronic structure, we considered the effect of self-energy corrections using a first-order calculation, e.g., $G^0W^0$. 
In this approximation the electron orbitals from the DFT are unmodified, but their energies are shifted by a complex-valued self-energy correction
\begin{align}
\vert \psi_{nk}^{GW} \rangle  &= \vert \psi_{nk}^{\text{DFT}} \rangle  \nonumber \\
E_{nk}^{GW} \;\,&= E_{nk}^\text{DFT} + \langle \psi_{nk}^{\text{DFT}}  \vert \Sigma(E_{nk}^{GW}) - V_{xc} \vert \psi_{nk}^{\text{DFT}}  \rangle \label{GW1} \\
E_{nk}^{GW} \;\,&= E_{nk}^\text{DFT} + \Delta^{GW}_{nk}  + i \Gamma^{GW}_{nk}  \nonumber 
\end{align}
where we divide the shift into its real part $\Delta$ and imaginary part $\Gamma$. 
Within this approximation the valence and conduction states are quasiparticles with spectral functions that follow a Lorentzian (Cauchy) distribution with a centroid of $E^\text{DFT} + \Delta$ and width of $\Gamma$.  
The $GW$ energy shift in Eq.~\ref{GW1} can be also be determined from considering the self-energy operator $\Sigma$ evaluated at the DFT energy. 

The self-energy corrections were calculated using {\sc abinit} \cite{abinit1}. 
A $4\times4\times4$ $\Gamma\!$-centered {\it k-}point grid was used, yielding 13 inequivalent points in the Brillouin zone. 
A scissors operator of 1.0~eV was applied before the screening and self-energy steps. 
The DFT orbitals were calculated with a 130~Ry plane-wave cut-off, and they were downsampled to 64~Ry and 56~Ry for the screening and self-energy calculation steps, respectively. 
The dielectric response was calculated from 0~eV to 49~eV with 1~eV intervals, and the self-energy was determined via contour integration \cite{PhysRevB.67.155208}. 
The dielectric matrix was calculated with a plane-wave cut-off of 20~Ry using 1466 bands. 
For the self-energy calculation 1000 bands were used and the exchange was truncated to 64~Ry.  
Corrections were determined for occupied (skipping the Li 1{\it s}) and 24 lowest unoccupied states.

\begin{figure}
\includegraphics[height=3.1in,trim=0 35 15 60,angle=270]{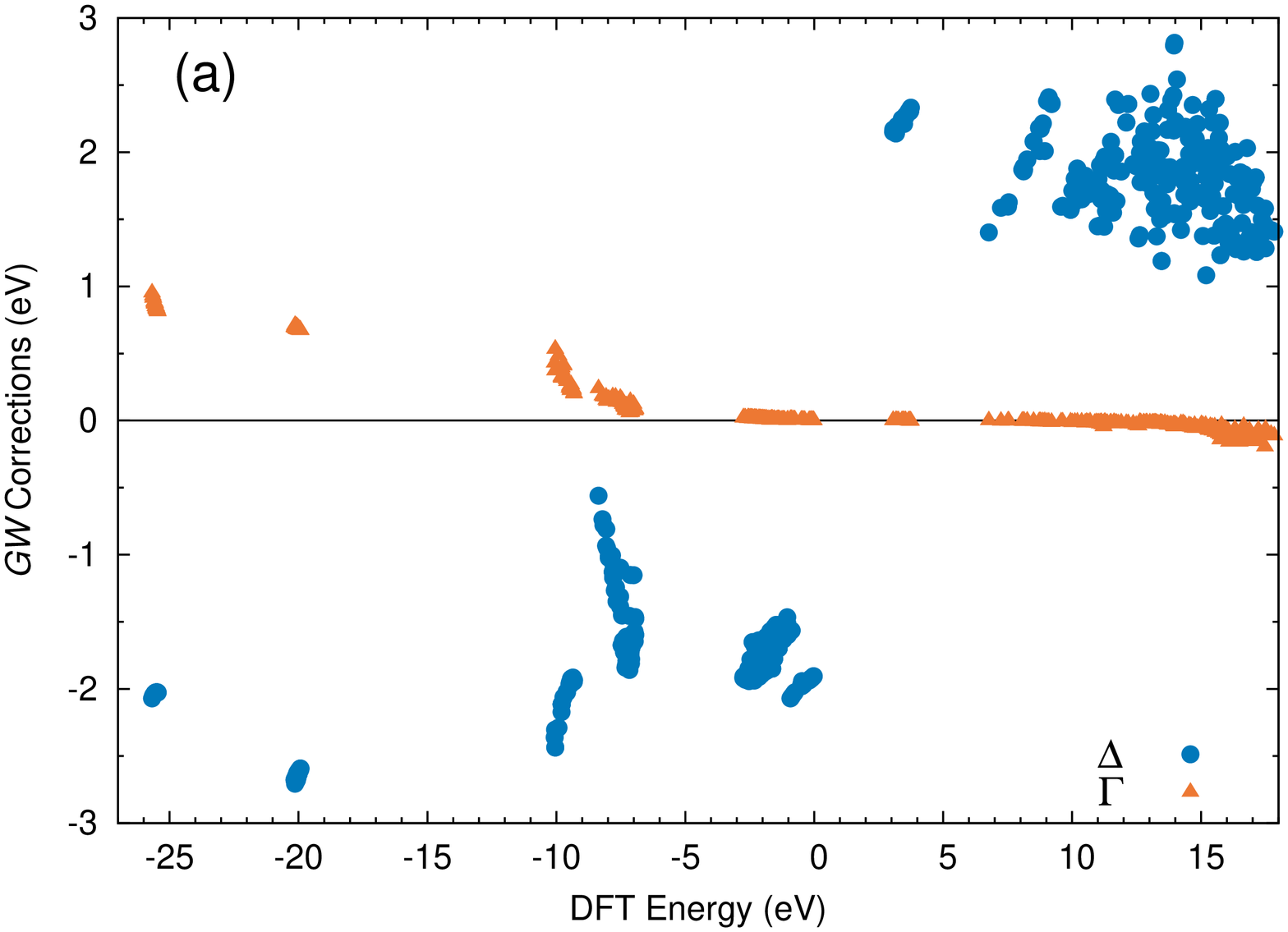}
\includegraphics[height=3.1in,trim=0 35 15 60,angle=270]{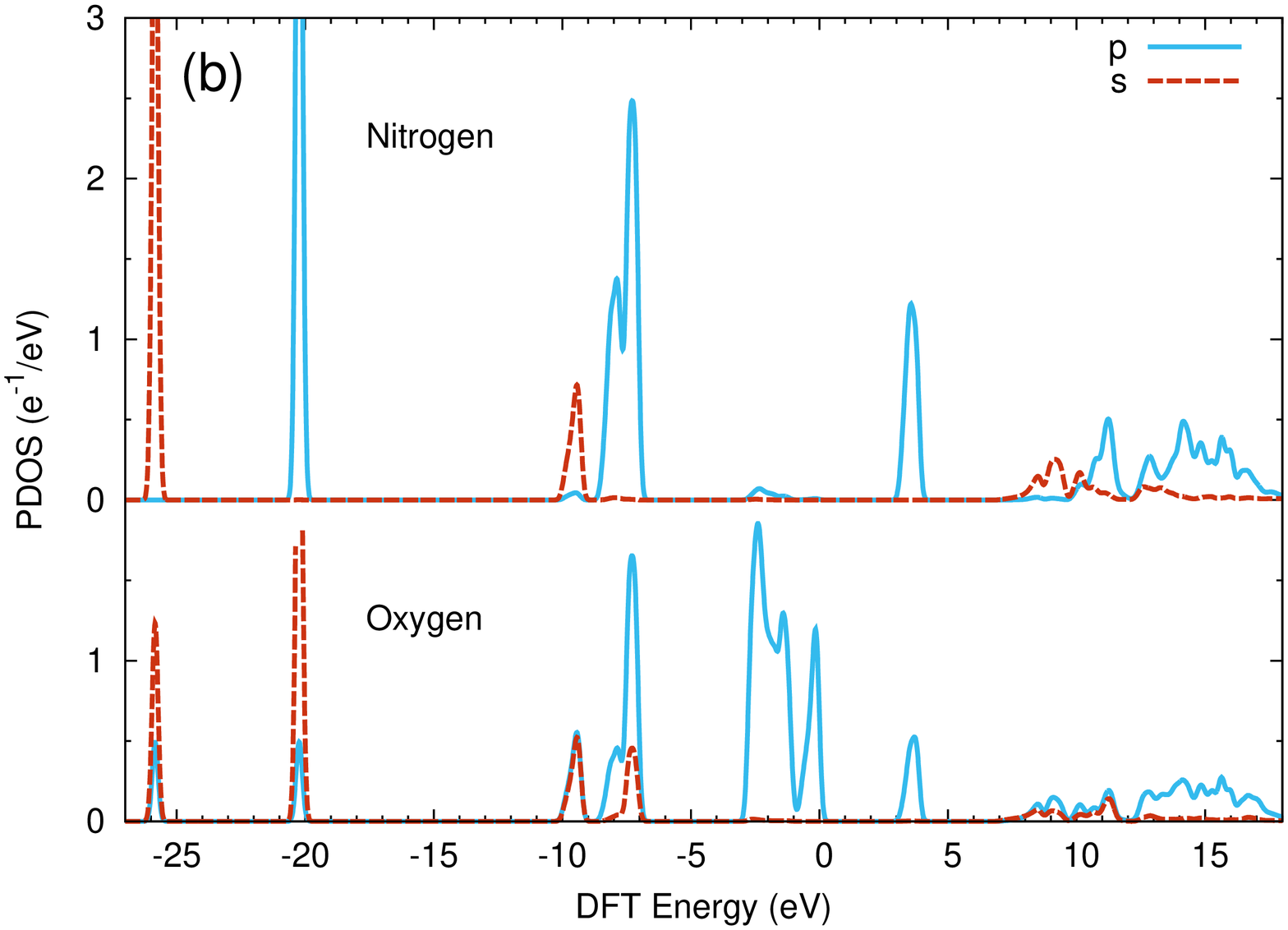}
\caption{Top (a), the real (blue circles) and imaginary (orange triangles) components of the $G^0W^0$ self-energy corrections (see Eq.~\ref{GW1}) as a function of the DFT energy relative to the valence band maximum. Bottom (b), the projected density of states (PDOS) around the nitrogen and oxygen atoms (per atom), artificially broadened by an 0.25~eV full-width at half-maximum Gaussian. The peaks below 20~eV are cropped. }
\label{gw_corrections}
\end{figure}

The real and imaginary components for the $GW$ corrections as a function of DFT energy are shown in Fig~\ref{gw_corrections}(a). 
For reference, the projected density of states (PDOS) is plotted below in Fig~\ref{gw_corrections}(b). 
Not pictured, the Li 1{\it s} states have a PBE energy of $-43.1$~eV with respect to the valence band maximum and no $GW$ corrections were calculated for them. 
 The nitrogen 2{\it s} states primarily make up the 1a$_1'$ orbitals which form a narrow band ranging from $-25.75$~eV to $-25.97$~eV and have a self-energy broadening of around 1~eV. 
 The four bands ranging from $-20.34$~eV to $-20.11$~eV are the 1e$'$ NO~$\sigma$ bonds. These NO~$\sigma$ orbitals have a calculated $GW$ broadening of 0.69~eV. 
The NO~$\pi$ orbitals ranging from $-9$~eV to $-7$~eV have an average broadening of 0.11~eV. 
All of the remaining valence states as well as the first 10~eV of unoccupied states show little broadening as is expected in an insulator.

\begin{figure}
\includegraphics[height=3.1in,trim=0 190 0 190,angle=270]{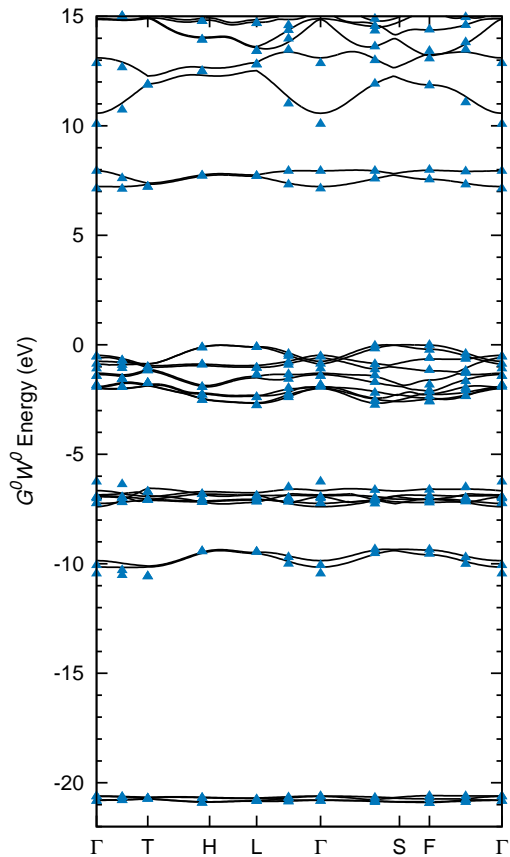}
\caption{The bandstructure of \LNO{} along several high-symmetry directions, following the path $\Gamma$--T--H$_2\vert$H$_0$--L--$\Gamma$--S$_0\vert$S$_2$--F--$\Gamma$. The energies shown are from a DFT calculation using the PBE exchange-correlation function, including a band-dependent approximate $GW$ correction. 
For reference the calculated $GW$ energies are shown by the blue triangles for select points in the Brillouin zone. The energies are relative to the $G^0W^0$ valence-band maximum.  }
\label{bandstructure}
\end{figure}

To plot the electronic band structure (Fig.~\ref{bandstructure}), we have followed the recommendation of Hinuma {\it et al.} \cite{HINUMA2017140} for the path through the Brillouin zone and special {\it k-}points (see Fig.~18 and Table~85 of Ref.~\onlinecite{HINUMA2017140}). 
The Li 1{\it s} and NO$_3$ 1a$_1'$  bands are not shown as they have almost no dispersion. 
The energies have been corrected using an approximate band-dependent $GW$ correction
\begin{align}
\Delta^{GW}_n = N^{-1} \sum_{k=1}^{N} \Delta^{GW}_{nk} 
\label{eq-gw}
\end{align}
which is the average of the corrections over the $N\!=4^3$ {\it k-}point grid on which we carried out the full $G^0W^0$ calculations. 
Except near the $\Gamma$-point, the band-average corrections are close to the {\it k-}dependent values. 
For the x-ray calculations, only the band-average corrections were applied. 

The PBE calculations of the optimized cell show that \LNO{} is an insulator with an indirect gap of 3.06~eV and a 3.41~eV direct gap. 
Our band-averaged {\it GW} calculations increase the fundamental gap by 4.16~eV to 7.22~eV. 
For the x-ray calculations (Sec.~III), spectra were generated by averaging over 10 disordered snapshots of a $2\times2\times2$ supercell. 
This disorder reduced the fundamental gap to an average of 6.1(2)~eV.

\section{X-ray Spectra}

\subsection{Experimental details}

X-ray absorption spectra (XAS) and X-ray emission spectra (XES) were obtained using the U49 beamline operated by the Physikalisch-Technische Bundesanstalt at the electron storage ring BESSY II. The beamline uses a plane grating monochromator and a spherical grating spectrometer. The samples consisted of powdered LiNO$_3$ pressed into In foil. A description of the monochromator and the spectrometer has been given previously in the literature \cite {SPIE2000, PRA.79.032503}. 

The plane grating monochromator was calibrated for energies around the N K edge by measuring the absorption spectrum of N$_2$ in a gas cell as detected by a photodiode for a given ratio of incidence and exit angles of the grating. The spectrometer was calibrated by quasi-elastic scattering of the monochromator beam by a target. The spectrometer was oriented in a scattering plane that was parallel to the polarization of the incident beam (p polarization). Because the scattering angle was 90$^{\circ}$, first-order elastic scattering was forbidden. A thin film \hBN{} target was used for the calibration, since quasi-elastic scattering is observed over a range of energies above the N K edge due to multiple or phonon-assisted scattering. A standard error not exceeding 100 meV was determined over the energy range covered in this work. 

The measurements were carried out at room temperature with the sample mounted at 45$^{\circ}$ from the incident beam and the takeoff angle of the spectrometer. Absorption spectra were obtained by monitoring the N K-fluorescence with a silicon drift detector at  45$^{\circ}$ to the scattering plane as well as recording the sample current.  Fluorescence intensities were corrected for self-absorption by taking a ratio of the two methods. X-ray emission spectra were accumulated at multiple positions on the sample for a total of 2400~s at each excitation energy. 

\subsection{Calculations}

%,trim=0 35 15 60
\begin{figure}
\includegraphics[height=3.1in,trim=0 35 15 60,angle=270]{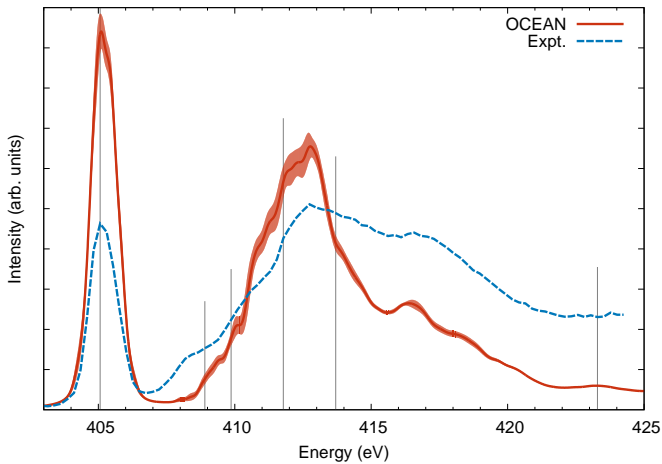}
\caption{The calculated (red, solid) and measured (blue, dashed) x-ray absorption spectra of LiNO$_3$. 
The width of the {\sc ocean} curve signifies the variation in the mean from using 10 snapshots. 
The light-grey lines indicate the energies at which RIXS data were taken (Fig.~5). }
\label{xas}
\end{figure}

Both x-ray absorption spectra (XAS) and RIXS were calculated using the Bethe-Sapeter equation (BSE) approach as implemented in the {\sc ocean} code \cite{ocean1,*ocean0,ocean2}. 
In order to capture the effects of vibrational disorder, $2\times2\times2$ supercells of \LNO{} were generated via the phonon-mode method detailed in \ref{phonons}. 
Each supercell contains 16 nitrogen sites and a total of 80 atoms. 
A $2^3$ {\it k-}point mesh and 304 conduction bands (496 total bands) were used for the BSE final states.
A single, $\Gamma$-centered {\it k-}point and 2208 bands were used to calculate the screening response to the core-hole potential. 
The electron wave functions were generated using the {\sc Quantum} ESPRESSO code \cite{espresso2,*espresso1,*espresso0} with the same pseudopotentials as for the earlier self-energy and phonon calculations. 
Band-averaged real-valued {\it GW} corrections were applied to the energies of the BSE states (Eq.~\ref{eq-gw}).
The conduction-band lifetimes [Fig.~\ref{gw_corrections}(a)] are small in the near-edge region, and only a constant core-hole lifetime broadening of 0.1~eV was applied to the absorption calculations. 
For the x-ray emission calculations, we note that the {\it p}-type nitrogen DOS can be easily divided, with states below $-15$~eV having an average lifetime broadening of 0.69~eV, while above that the broadening is no more than 0.11~eV.  
We therefore applied separate constant broadening to the spectra, once again neglecting the small contributions from the conduction bands. 
The generalized minimal residual method (GMRES) \cite{GMRES} was used to generate the RIXS intermediate states, while the absorption and emission spectra were generated using the Haydock method \cite{Haydock}. 
An isotropic dielectric constant of $\epsilon_0=3.0$ was used as input to the model dielectric response used in both the core- and valence-level BSE, from an index of refraction of 1.735 \cite{Patnaik}. 
Further details of the use of GMRES and how RIXS is calculated using {\sc ocean} are included in Ref.~\onlinecite{PhysRevB.94.035163}.

The electron-photon interactions were treated within the dipole limit, and, to simulate the powdered sample used in the experiment, several polarization directions for both incoming and outgoing x-ray photons were used, subject to the experimentally-determined constraint that the outgoing photon momentum was aligned with the incoming photon polarization. 
The presented spectra are an average over nitrogen sites, photon polarizations, and supercells. 
For each calculated spectrum shown, the thickness of the curve denotes the variance of the mean, or  statistical uncertainty from averaging over only a finite number of supercells. 
An additional, Gaussian broadening of 0.25~eV (0.50~eV) was applied to the calculated absorption (emission) spectra before averaging. 
The alignment of the XAS energy scale of the {\sc ocean} calculations follows previous work \cite{PhysRevB.94.035163}.
The emission energy of the calculations is fixed relative to the absorption alignment, including the band-gap correction from the {\it GW}.

\subsection{X-ray absorption}

We present the measured and calculated XAS in Fig.~\ref{xas}. 
The overall structure is common to K edges of {\it s-p} systems with a trigonal or hexagonal geometry (locally D$_{3h}$), such as boron in BF$_3$ or hexagonal boron nitride, carbon in various carbonates CO$_3^{2-}$ or graphite, and other nitrates. 
The dipole allowed transitions (1{\it s}$\rightarrow$n{\it p})  are split in energy with a sharp exciton-well separated from a broader feature. 
The {\it p}$_z$ states (405~eV) have little overlap with the bonding orbitals and are significantly lower in energy than the {\it p}$_\textit{xy}$ excitations (410~eV to 415~eV). 
This separation is also evident in the band structure (Fig.~\ref{bandstructure}).

Overall we find good agreement between the measured spectrum and the calculations with respect to the width of the main exciton and the spacing between the main exciton and the features near 413~eV and 417~eV. 
There is an overestimation of the strength of the main exciton and corresponding underestimation of spectral weight at higher energies, but this could also point to an insufficient treatment of self-absorption effects in the measured spectrum which are expected to vary strongly with energy. 
We find disagreement in the onset of absorption after the main exciton, with the experimental spectrum showing an onset near 408~eV, about 1~eV lower in energy than the calculation. 
While this difference could be the result of deficiencies in the calculated electronic band structure, we believe that the PBE+{\it GW} calculations should be more accurate than this for \LNO{}. 
Instead we suggest that some degree of atomic disorder is present in the experimental samples but absent in our model, such as distortions outside of the harmonic approximation in which the phonons were calculated.

\begin{figure}
\includegraphics[width=3.1in,trim=60 0 60 0,angle=0]{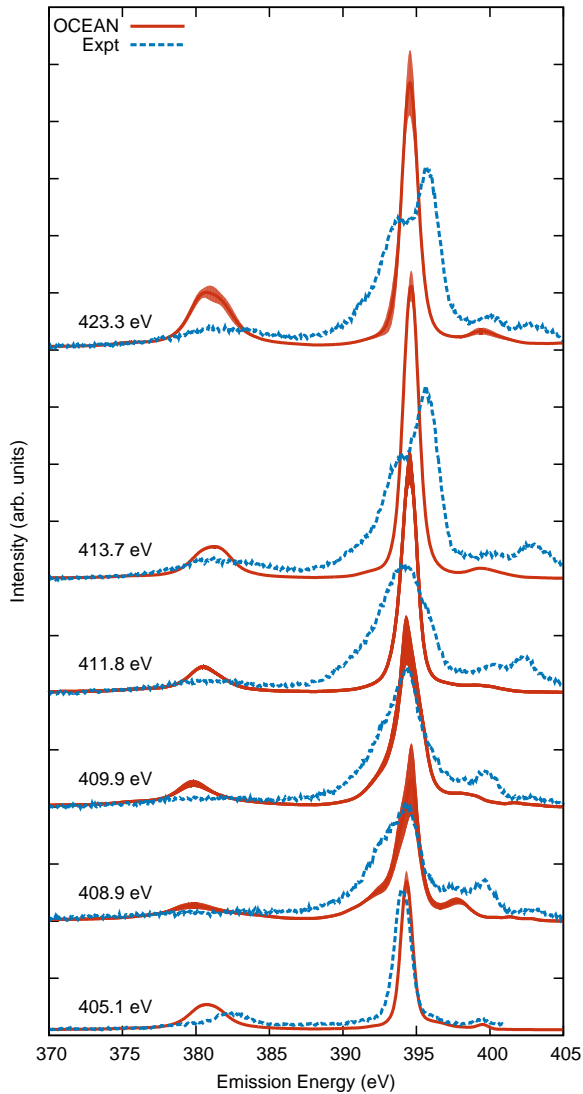}
\caption{The calculated (red, solid) and measured (blue, dashed) resonant x-ray inelastic scattering spectra of LiNO$_3$ at the nitrogen K edge. 
The width of the {\sc ocean} curve signifies the variation in the mean from using 10 snapshots. }
\label{rixs}
\end{figure}

\subsection{RIXS}

In Fig.~\ref{rixs} we present measured and calculated RIXS spectra, labeled by the incoming x-ray energy. 
For reference these energies are also marked on the plot of the x-ray absorption (Fig.~\ref{xas}). 
Measurements were taken at four additional absorption energies between 414~eV and 424~eV, but little change was observed in the x-ray emission spectra. 
The measured emission for 405.1~eV incident x-rays is truncated above 401~eV to remove the tail from the elastic peak. 
Overall, the changes in the emission with incident energy are minor, and the changes in the calculated emission spectra are even more so. 
We note that in the alignment of the absorption and emission calculations with the measured spectra there is only a single free parameter which was chosen to match the position on the main exciton in the absorption at 405~eV. 
The subsequent position of the calculated emission peaks was then fixed by the band structure, increased by the {\it GW} corrections, and finally decreased by the calculated exciton binding for both the core-level and valence-level excitons within the BSE. 

Starting at resonance, we find good agreement between the measured and calculated RIXS for the 405.1~eV incident energy beam. 
The NO~$\sigma$ peak near 382~eV is over-bound by 1.5~eV in the calculation, while the NO~$\pi$ peak is under-bound by less than 1~eV. 
However, the relative weights of the two main features, along with the more delocalized valence band feature at 399~eV agree remarkably well. 
Furthermore, the widths of the peaks are modeled well by the calculation. 
Using a displaced Gaussian and constant background, we are able to determine the width of the calculated and measured NO~$\sigma$ peak. 
Without the {\it GW} or vibrational effects, the calculated NO~$\sigma$ peak is very narrow, with almost no dispersion from the band structure (Fig.~\ref{bandstructure}). 
Vibrational disorder alone increases the full width at half maximum to 2.2~eV, while the inclusion of the imaginary component of the {\it GW} corrections increases it further to 3.3~eV.  
This is 10~\% larger than the measured full width half maximum of 3.0~eV. 
Without considering broadening from the imaginary part of the self-energy, the difference between the measured and calculated width of the $\sigma$ peak would incorrectly point to additional atomic disorder or dispersion deep in the valence band. 
This situation highlights the importance of using a complex many-body self-energy as a foundation for spectroscopy calculations. 

As the incident photon energy is tuned above the main exciton, the observed emission energy changes. 
The $\pi$ features around 394~eV first broaden and then split into two distinct peaks. 
The $\sigma$ feature at 382~eV also broadens, but any split is obscured by the self-energy broadening. 
At incident energies of 408.9~eV and 409.9~eV, the suppression of the $\sigma$ peak and asymmetry in the the $\pi$ peak are captured in our calculation.
However, at higher energies no splitting of the $\pi$ peak or additional broadening of the $\sigma$ peak are observed in the calculations. 
We hypothesize that this splitting is an excited-state effect, driven by atomic relaxation during the RIXS process, {\it i.e.}, Jahn-Teller distortion. 
The local potential energy surface can change dramatically in the presence of the x-ray absorption exciton, driving changes in the local atomic configuration. 
This effect is well-known, and has been studied computationally in aqueous systems \cite{C5CP04898B}.

\section{Conclusions}

We presented measured XAS and RIXS spectra of \LNO{} along with {\it GW}-BSE calculations of the same using the {\sc ocean} code. 
The x-ray emission spectra show a sizable broadening of the NO~$\sigma$ peak, which was previously noted in the x-ray emission attributed to the  NO$_3^-$ ion within ammonium nitrate \cite{PhysRevB.90.205207}. 
By including broadening from both the ground-state vibrational disorder and the imaginary part of the {\it GW} self-energy corrections, we are able to match the observed width of the NO~$\sigma$ peak. 
This substantial, peak-dependent broadening is a general effect in NO$_3^-$ compounds. 
In addition to probing the real-valued components of the electronic structure, we show that RIXS spectroscopy can also measure broadening from the valence-band lifetime.
In this way RIXS provides a measurement of the full complex self-energy and may serve as a unique check on the accuracy of this aspect of first-principles self-energy calculations. 

Challenges remain in incorporating the effects of the atoms in the crystal as dynamic quantum particles instead of a fixed lattice of point charges. 
Our phonon mode method adequately captures modifications of the x-ray absorption spectrum as a result of the intrinsic disorder from vibrations. 
However, a more sophisticated approach seems necessary to accurately simulate resonant emission in extended systems. 
The deficiency may be especially notable in cases such as \LNO{} where the system has localized or molecular characteristics (the multi-atom NO$_3^{-}$ ion) where the atoms are in a strongly anisotropic binding environment. 
Methods designed for embedded localized or molecular systems may be applicable to molecular solids, but it would also be of interest to develop  methods that are able to approach this problem from an extended, delocalized picture. 

\medskip

Certain software packages are identified in this paper to foster understanding. Such identification does not imply recommendation or endorsement by the National Institute of Standards and Technology, nor does it imply that these are necessarily the best available for the purpose.

%\bibliography

\end{document}